\documentclass[default]{sn-jnl}

\usepackage{graphicx}%
\usepackage{multirow}%
\usepackage{amsmath,amssymb,amsfonts}%
\usepackage{amsthm}%
\usepackage{mathrsfs}%
\usepackage[title]{appendix}%
\usepackage{xcolor}%
\usepackage{textcomp}%
\usepackage{manyfoot}%
\usepackage{booktabs}%
\usepackage{algorithm}%
\usepackage{algorithmicx}%
\usepackage{algpseudocode}%
\usepackage{listings}%
\usepackage{parskip}

\theoremstyle{thmstyleone}%
\theoremstyle{thmstyletwo}%

\theoremstyle{thmstylethree}%

\begin{document}

\title[Article Title]{Object Recognition in Human Computer Interaction:- A Comparative Analysis}


\author*[1]{\fnm{Kaushik} \sur{Ranade}}\email{kaushikranade.dev@gmail.com}

\author[1]{\fnm{Tanmay} \sur{Khule}}\email{khule.tanmay.dev@gmail.com}

\author[2]{\fnm{Riddhi} \sur{More}}\email{riddhi.more1@ontariotechu.net}
\abstract{Human-computer interaction (HCI) has been a widely researched area for many years, with continuous advancements in technology leading to the development of new techniques that change the way we interact with computers. With the recent advent of powerful computers, we recognize human actions and interact accordingly, thus revolutionizing the way we interact with computers. The purpose of this paper is to provide a comparative analysis of various algorithms used for recognizing user faces and gestures in the context of computer vision and HCI. This study aims to explore and evaluate the performance of different algorithms in terms of accuracy, robustness, and efficiency. This study aims to provide a comprehensive analysis of algorithms for face and gesture recognition in the context of computer vision and HCI, with the goal of improving the design and development of interactive systems that are more intuitive, efficient, and user-friendly.}




\maketitle

\section{Introduction}\label{sec1}

The field of human-computer interaction (HCI) has undergone significant transformation in recent years with  advancements in computing power, computer vision, and machine learning techniques. The demand for more interactive and natural user interfaces has led to the development of systems that can recognize human actions and gestures \cite{Surya_Devarajan2014-nh}, \cite{beddiar2020vision}, \cite{choi2008view}. These smart-systems enable users to interact with computers more effectively and intuitively, making HCI a topic of extensive research in academia and industry. One critical aspect of HCI is the recognition of users' faces and gestures. Recognizing faces and gestures accurately and in real-time is an essential component of many applications, including gaming, security, and human-robot interaction. Several approaches have been proposed in the literature for face and gesture recognition, including traditional computer vision-based methods \cite{CutlerTurk98}, \cite{hasanuzzaman2004real}, and more recent deep learning-based techniques \cite{hu2020deep}.

Face recognition verifies identity via facial features, with practical applications in security, biometrics, and HCI. It has been an area of extensive research for decades, with a wide range of practical applications, including security and surveillance systems, biometric identification, and access control. Similar to face recognition, gesture recognition is an important area of research in human-computer interaction, allowing machines to recognize and understand human actions and gestures. It has numerous practical applications, including gaming, virtual reality, and human-robot interaction. In recent years, there has been significant progress in the development of face and gesture recognition techniques, driven largely by advances in machine learning and computer vision algorithms. These techniques include traditional computer vision-based methods and more recent deep learning-based approaches. In this paper, we will look at them in detail.

\section{Methodology}\label{sec2}

Our proposed human-computer interaction (HCI) system consists of two main phases: authentication and continuous tracking. To ensure system security, we leverage advanced face recognition algorithms in the authentication phase to authenticate users. Once the user is verified, the system enters the second phase of continuous tracking, where gesture recognition technology is utilized to track and interpret the user's hand gestures in real-time. This approach provides a more natural and intuitive way for users to interact with the system, reducing the reliance on traditional input methods such as a keyboard or mouse. By making use of state-of-the-art computer vision and deep learning algorithms, we aim to develop an efficient and accurate HCI system that is both user-friendly and secure. Our proposed approach guarantees that the system is controlled by an authorized person, ensuring the privacy and security of the user's data and interactions.
\subsection{Algoirthms}
Face recognition is an essential component of any modern HCI system, allowing machines to identify and verify the identity of individuals by analyzing facial features. In recent years, there has been significant progress in the development of face recognition techniques, driven largely by advances in deep learning and computer vision algorithms. In this paper, we will provide an in-depth comparison between the following algorithms: Eigen Faces \cite{turk1991face}, Viola-Jones Algorithm \cite{viola2001rapid}, Hog cascade + CNN \cite{hung2021face}, and Key Point-based methods. We also extend our research into the next phase of our proposed HCI system, investigating various gesture recognition techniques, such as bit image clustering \cite{panwar2012hand}, convolutional neural networks \cite{pinto2019static} and key point-based methods \cite{schneider2019gesture}.
\subsection{Dataset}
\subsubsection{Face Recognition}
For comparing and analyzing different algorithms, a good dataset is essential. Any bias or incorrect data can change the outcome of our analysis. For face recognition, we chose The Labeled Faces in Wild dataset \cite{learned2016labeled}. This dataset is a widely used benchmark for evaluating face recognition algorithms in real-world scenarios. It contains over 13,000 images of faces collected from the internet, with significant variations in pose, lighting, and expression. The dataset has been instrumental in advancing the state-of-the-art in face recognition, and many of the top-performing algorithms have been evaluated on this dataset.
\subsubsection{Gesture Recognition}
For the analysis of gesture recognition algorithms, we used the ASL dataset by the SigNN team \cite{noauthor_undated-iz}. The ASL Sign Language Pictures dataset is a collection of over 8,000 images of American Sign Language (ASL) hand signs representing letters A to I and K to Y, with the letters J and Z excluded. The images were collected using a digital camera and featured a diverse range of hand shapes, orientations, and lighting conditions. This dataset can be used for training machine learning models to recognize and interpret ASL hand signs, enabling improved communication and accessibility for the deaf and hard-of-hearing community. We use this dataset to compare and contrast various gesture recognition techniques because it provides a wide variety of static signs.

\section{Review and Comparison of various Algorithms}\label{3}
In this section, we present a comparative analysis of various face and gesture recognition algorithms in terms of robustness and efficiency. Our study aims to identify the most suitable algorithm for the proposed human-computer interaction (HCI) system.
\subsection{Face Recognition Algorithms}
Face recognition algorithms are primarily classified in three ways: local approaches, holistic approaches, and hybrid approaches. We evaluated four different face recognition algorithms: Eigen Faces, Viola-Jones Algorithm, Hog cascade + CNN, and Key Point-based methods. Each algorithm was tested on the Labeled Faces in Wild dataset \cite{learned2016labeled}, which consists of over 13,000 images of faces collected from the internet.

Eigen Faces – Eigenfaces is the most primitive algorithm for face recognition, which works by projecting an image on a lower dimensional feature space to extract the compact representations of the face. Then, the algorithms compare the projection of the given image with projections of a set of training images known to the classifier and find the closest match \cite{turk1991face}. During recognition, an input image is projected onto the same space, and the closest match is found based on Euclidean distance. Eigenfaces exploit the fact that faces in photographs are usually upright in orientation and hence work best with such pictures but have limitations in handling complex variations in pose, lighting, and expression.

Viola-Jones algorithm- The Viola-Jones algorithm is a real-time face detection algorithm that uses haar-like features and the concept of cascaded classifiers to detect faces. The algorithm works by detecting edge features or corners, and differences in contrasts for various regions in the image, then trains the cascaded classifiers to detect face-likely regions in an input image and discards the background. The algorithm works best for real-time systems but may suffer from false positives and does not perform well if the image is partially occluded.\cite{viola2001rapid}

Hog cascade + CNN - Hog cascade + CNN is a hybrid approach that combines traditional computer vision techniques with deep learning. The algorithm uses a Haar cascade classifier [viola jones ka ref] to detect faces and then feeds the detected regions into a convolutional neural network (CNN) for further processing. This approach is suitable for handling complex variations in pose, lighting, and expression and can achieve high accuracy on large and complex datasets. However, it suffers from slow processing times and may require significant computational resources.\cite{felzenszwalb2009object}

Key Point-based methods: Key Point-based methods use local features, such as Scale-Invariant Feature Transform (SIFT), Speeded Up Robust Feature (SURF), or Oriented FAST and Rotated BRIEF (ORB), to detect and describe distinctive key points in an image. These methods use a set of landmark points on the face, such as the eyes, nose, and mouth, to extract local features for face recognition. The algorithm then matches the key points of the test image with the key points of a set of training images to find the closest match. Key point-based methods are robust to variations in pose, lighting, and expression and can handle occlusion and partial face occlusion. However, they may suffer from the problem of feature-matching ambiguity, especially when the points are sparse.

In summary, each of these face recognition algorithms has its own strengths and weaknesses, and the choice of algorithm depends on the specific application requirements. Eigenfaces are fast and computationally efficient but can suffer from variations in lighting, pose, and facial expressions. The Viola-Jones algorithm is computationally efficient and can detect faces in real-time, but it can suffer from false positives and misses faces in certain orientations and lighting conditions. HOG + CNN algorithms are highly accurate and robust to variations in lighting, pose, and facial expressions, but they can be computationally expensive. Key point-based methods are highly accurate and robust to variations in lighting and pose, but they can be sensitive to occlusions and facial expressions. For our case, we simply needed a highly accurate, robust, and pose- and occlusion-invariant algorithm; hence, using CNN or deep learning methods was the best choice. Many popular computer vision APIs have already implemented various deep-learning algorithms for facial recognition. Additionally, in the literature, it is evident that deep learning-based methods outperform other methods on most datasets \cite{wang2021deep}.
\subsection{Deep Learning-Based Techniques}
Deep learning-based techniques for hand gesture recognition have shown significant improvements over traditional computer vision-based methods, especially in complex scenarios with varying lighting conditions and background clutter. These techniques involve training deep neural networks using large amounts of data and require high computational resources. We explored the performance of three deep learning algorithms: neural network, CNN, and LSTM.

\begin{figure}[h]%
\centering
\includegraphics[width=0.9\textwidth]{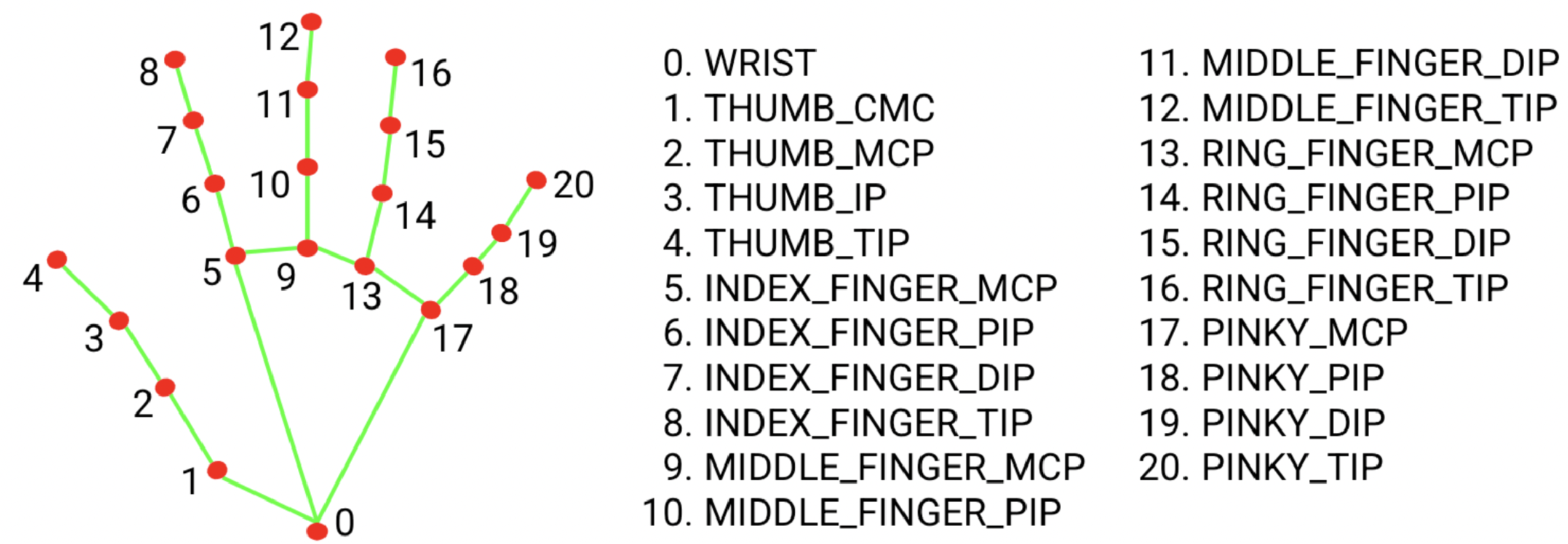}
\caption{ This image \cite{mediapipe} shows the 21 hand landmarks that are detected by the MediaPipe Hand Landmark detection API. The landmarks, numbered and labeled in the image, are utilized to track the position and orientation of the hand in real-time.}\label{fig1}
\end{figure}
First, we isolate image frames from the real-time video input and extract hand key points using a media pipe hand landmark model. This model achieves highly accurate localization of 21 3D hand-knuckle coordinates within detected hand regions through direct coordinate prediction. This approach allows the model to learn a reliable internal representation of hand pose, which enables it to handle challenging scenarios such as partially visible hands and self-occlusions with robustness. Then, we trained each of these algorithms on these key points. During inference, we used a pre-trained hand detection model and passed on the key points to save versions of our network.

Convolutional Neural Networks or (CNN) is a type of deep learning algorithm that has shown remarkable performance in image recognition tasks, including hand gesture recognition. CNNs learn to extract hierarchical features from the input images by applying a series of convolutional filters. These filters convolve over the input image and extract features such as edges, corners, and other patterns. This hierarchical feature extraction allows CNNs to learn high-level features that are more relevant to the hand gesture recognition task. In our experiments, we used a CNN architecture consisting of multiple convolutional layers, followed by pooling layers and fully connected layers.
We trained the network on the ASL Sign Language Pictures dataset using the key points extracted from the hand region. Our results showed that CNNs had the highest accuracy among all the tested algorithms, but they required a large amount of training data and computational resources. Furthermore, CNNs were unable to generalize on unseen or new data due to poor regularization of input data, which leads to overfitting, as seen in a few iterations, and makes CNNs less robust.

LSTMs on the other hand are a type of recurrent neural network (RNN) that can model temporal dependencies in sequential data. LSTMs can learn and remember long-term dependencies between the input sequences and are particularly effective for recognizing dynamic hand gestures that involve motion. 

\begin{figure}[h]%
\centering
\includegraphics[width=0.3\textwidth]{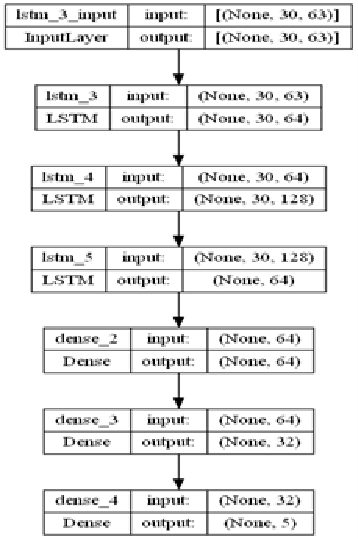}
\caption{ This image illustrates the architecture of LSTM network}\label{fig2}
\end{figure}

Our LSTM architecture consisted of an input layer of dimension (30, 63) followed by two LSTM layers, each of size 128 and 64. We then used a series of three dense layers of size 64, 32, and the last one of size equal to output classes 5. However, in our experiments, we focused on recognizing static hand signs, and LSTMs were not performing up to mark for this task. Nonetheless, LSTMs show great promise for future applications in recognizing dynamic hand gestures.

Support vector machine (SVM) is a traditional machine learning algorithm that has been widely used for hand gesture recognition tasks. SVM works by finding the hyperplane that maximally separates the data points into different classes. In the context of hand gesture recognition, SVM is trained on a set of hand-crafted features extracted from the input images, such as histograms of oriented gradients (HOG)\cite{miron2019hand} or local binary patterns (LBP)\cite{lahiani2019hand}. During testing, the SVM classifier takes in the extracted features of a new image and assigns it to one of the predefined classes.

SVMs have shown good performance in hand gesture recognition tasks, especially in scenarios with limited training data. SVM is also relatively computationally efficient and has low memory requirements, making it suitable for real-time applications. However, similar to other traditional computer vision-based methods, SVM may not perform well in complex scenarios with varying lighting conditions, background clutter, and occlusions. In comparison to deep learning-based methods, SVM may also have limited ability to learn high-level features automatically and generalize to unseen data.

In our experiments, we used key points detected from the input images as features for SVM classification. While the performance of SVM was not as good as the deep learning-based methods we tested, it still achieved reasonable accuracy and can be a viable option for certain applications with limited training data and computational resources.

Neural networks (NNs) are a type of deep learning model that can be used for classification tasks. NNs consist of several layers of interconnected nodes that process and transform the input data. These layers include input layers, hidden layers, and output layers. In hand gesture recognition, NNs are trained on a dataset of hand gestures and their corresponding labels. During inference, the model receives an input image and produces a label that corresponds to the recognized gesture.

Our NN architecture consists of an input layer of dimension 21*3 following a dropout of 0.2, then a dense layer of size 40 followed by another dropout of 0.4, then a dense layer of size 20, and finally another dense layer of size equal to the number of classes.

\begin{figure}[h]%
\centering
\includegraphics[width=0.4\textwidth]{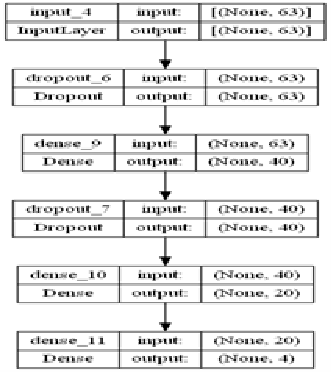}
\caption{ This image illustrates the Neural Network architecture used to predict the gestures.}\label{fig3}
\end{figure}
Finally, to compare the performance of different hand gesture recognition algorithms, we evaluated their performance based on accuracy, robustness, and efficiency.
\begin{table}[h]
\label{table:models}
\begin{tabular}{cccc}
\toprule
\textbf{Model} & \multicolumn{3}{c}{\textbf{Parameters}} \\
\cmidrule(lr){2-4}
 & \textbf{Robustness} & \textbf{Efficiency} & \textbf{Accuracy} \\
\midrule
Key point + SVM & Low & Medium & 70\% \\
Key point + NN & High & High & 75\% \\
CNN & Low & High & 90\% \\
LSTM & High & Low & 85\% \\
\bottomrule
\end{tabular}
\caption{Comparison of different models.\footnotesize (The accuracy values are estimated on preliminary findings).}
\end{table}
\\
\\
\section{Conclusion}\label{sec13}

Different face and gesture recognition algorithms were evaluated for their robustness and efficiency. For face recognition, Eigenfaces, Viola-Jones Algorithm, HOG + CNN, and Key Point-based methods were studied. Each algorithm has its advantages and limitations. Deep learning-based methods such as CNNs are suitable for highly accurate and robust face recognition. Traditional computer vision-based techniques such as BIC and KPM are computationally efficient and require less training data, but may not perform well in complex scenarios. For gesture recognition, deep learning-based methods such as CNN, ANN, and LSTM are suitable for complex scenarios with varying lighting conditions, background clutter, and occlusions, while traditional computer vision-based techniques such as BIC and KPM are suitable for real-time processing and simple scenarios.

The choice of algorithm depends on the specific requirements of the application. Given that the focus of our application is on real-time processing, we decided to choose key points with  a neural network approach because it struck a balance between computational efficiency and algorithmic complexity.

\section*{Declarations}
\subsection{Funding}
No funding was received to assist with the preparation of this manuscript or for conducting this study.
\subsection{Competing Interests}
The author doesn't have relevant financial or non-financial interests to disclose. Rather no competing interests to declare that are relevant to the content of this article.
\bmhead{Acknowledgement}
We would like to express our gratitude to Professor Ranjana Singh for her valuable guidance and support throughout this project.
\newpage
\bibliographystyle{unsrt}
\bibliography{Article.bib}
\end{document}